\documentclass[twocolumn]{aastex631}
\usepackage{newtxtext,amsfonts}

\usepackage[T1]{fontenc}

\DeclareRobustCommand{\VAN}[3]{#2}
\let\VANthebibliography\thebibliography
\def\thebibliography{\DeclareRobustCommand{\VAN}[3]{##3}\VANthebibliography}

\usepackage{graphicx}	
\usepackage{amsmath}	
\usepackage{float}
\usepackage[autopunct=true]{csquotes}

\begin{document}
\newcommand{\pfh}[1]{\textcolor{red}{#1}}

\newcommand{\GIZMO}{{\small GIZMO }}
\newcommand{\SF}{{\small STARFORGE }}

\newcommand{\gizmourl}{\href{http://www.tapir.caltech.edu/~phopkins/Site/GIZMO.html}{\url{http://www.tapir.caltech.edu/~phopkins/Site/GIZMO.html}}}

\newcommand{\datastatement}[1]{\begin{small}\section*{Data Availability Statement}\end{small}{\noindent #1}\vspace{5pt}}
\newcommand{\microGauss}{\mu{\rm G}}
\newcommand{\Bangle}{\theta_{B}}
\newcommand{\Alf}{{Alfv\'en}}
\newcommand{\BV}{Brunt-V\"ais\"al\"a}
\newcommand{\fref}[1]{Fig.~\ref{#1}}
\newcommand{\sref}[1]{\S~\ref{#1}}
\newcommand{\aref}[1]{App.~\ref{#1}}
\newcommand{\tref}[1]{Table~\ref{#1}}

\newcommand{\Dt}[1]{\frac{\mathrm{d} #1}{\mathrm{dt}}}
\newcommand{\initvalupper}[1]{#1^{0}}
\newcommand{\initvallower}[1]{#1_{0}}
\newcommand{\driftvel}{{\bf w}_{s}}
\newcommand{\driftvelmag}{w_{s}}
\newcommand{\driftvelhat}{\hat{{\bf w}}_{s}}
\newcommand{\driftveli}[1]{{\bf w}_{s,\,#1}}
\newcommand{\driftvelmagi}[1]{w_{s,\,#1}}
\newcommand{\dustvel}{{\bf v}_{d}}
\newcommand{\gasvel}{{\bf u}_{g}}
\newcommand{\gasden}{\rho_{g}}
\newcommand{\rhobase}{\rho_{\rm base}}
\newcommand{\gaspressure}{P}
\newcommand{\dustden}{\rho_{d}}
\newcommand{\rhodust}{\dustden}
\newcommand{\rhogas}{\gasden}
\newcommand{\resolution}{\Delta x_{0}}
\newcommand{\opticaldepth}{\tau_{\rm ext}}
\newcommand{\tauparam}{\tau_{\rm SL}}
\newcommand{\ts}{t_{s}}
\newcommand{\cs}{c_{s}}
\newcommand{\vA}{v_{A}}
\newcommand{\tL}{t_{L}}
\newcommand{\grainsuff}{_{\rm grain}}
\newcommand{\internaldensity}{\bar{\rho}\grainsuff^{\,i}}
\newcommand{\grainsize}{\epsilon\grainsuff}
\newcommand{\grainsizebar}{\bar{\epsilon}\grainsuff}
\newcommand{\grainmass}{m\grainsuff}
\newcommand{\graincharge}{q\grainsuff}
\newcommand{\grainchargeZ}{Z\grainsuff}
\newcommand{\grainsizemax}{\grainsize^{\rm max}}
\newcommand{\grainsizemin}{\grainsize^{\rm min}}
\newcommand{\B}{{\bf B}}
\newcommand{\Bmag}{|\B|}
\newcommand{\Bhat}{\hat\B}
\newcommand{\bhat}{\Bhat}
\newcommand{\acc}{{\bf a}}
\newcommand{\Lbox}{L_{\rm box}}
\newcommand{\Lscale}{H_{\rm gas}}
\newcommand{\sizeparam}{\tilde{\alpha}}
\newcommand{\sizeparammax}{\sizeparam_{\rm m}}
\newcommand{\chargeparam}{\tilde{\phi}}
\newcommand{\chargeparammax}{\chargeparam_{\rm m}}
\newcommand{\accparam}{\tilde{a}_{\rm d}}
\newcommand{\accparammax}{\tilde{a}_{\rm d,m}}
\newcommand{\accabsmax}{{a}_{\rm d,m}}
\newcommand{\accsizedep}{\psi_{a}}
\newcommand{\gravparam}{\tilde{g}}
\newcommand{\dustgas}{\mu^{\rm dg}}
\newcommand{\dustgashat}{\hat{\mu}^{\rm dg}}
\newcommand{\angstrom}{\mbox{\normalfont\AA}}

\def\lesssimA#1#2{\mathrel{\vcenter{\offinterlineskip%
    \ialign{\hfil##\hfil\cr$#1<$\cr$#1\sim$\cr}%
}}}

\def\lesssim{\mathpalette\lesssimA{}}

\def\app#1#2{%
  \mathrel{%
    \setbox0=\hbox{$#1\sim$}%
    \setbox2=\hbox{%
      \rlap{\hbox{$#1\propto$}}%
      \lower1.1\ht0\box0%
    }%
    \raise0.25\ht2\box2%
  }%
}
\def\approxprop{\mathpalette\app\relax}

\newcommand{\nadine}[1]{\textcolor{red}{#1}}
\title{Thermodynamics of Giant Molecular Clouds: The Effects of Dust Grain Size}
\shorttitle{STARFORGE: Dust in GMCs}

\correspondingauthor{Nadine H.~Soliman}
\email{nsoliman@caltech.edu}
\author[0000-0002-6810-1110]{Nadine H.~Soliman}
\affiliation{TAPIR, Mailcode 350-17, California Institute of Technology, Pasadena, CA 91125, USA}

\author{Philip F.~Hopkins}
\affiliation{TAPIR, Mailcode 350-17, California Institute of Technology, Pasadena, CA 91125, USA}

\author{Michael Y. Grudi\'{c}}
\affiliation{Carnegie Observatories, 813 Santa Barbara St, Pasadena, CA 91101, USA}
\shortauthors{Soliman et al.}



\begin{abstract}
The dust grain size distribution (GSD) likely varies significantly across star-forming environments in the Universe, but its impact on star formation remains unclear. This ambiguity arises because the GSD interacts non-linearly with processes like heating, cooling, radiation, and chemistry, which have competing effects and varying environmental dependencies. Processes such as grain coagulation, expected to be efficient in dense star-forming regions, reduce the abundance of small grains and increase that of larger grains. Motivated by this, we investigate the effects of similar GSD variations on the thermochemistry and evolution of giant molecular clouds (GMCs) using magnetohydrodynamic simulations spanning a range of cloud masses and grain sizes, which explicitly incorporate the dynamics of dust grains within the full-physics framework of the \SF project.  We find that grain size variations significantly alter GMC thermochemistry: with the leading-order effect is that larger grains, under fixed dust mass, GSD dynamic range, and dust-to-gas ratio, result in lower dust opacities. This reduced opacity permits ISRF and internal radiation photons to penetrate more deeply. This leads to rapid gas heating and inhibited star formation. Star formation efficiency is highly sensitive to grain size, with an order of magnitude reduction when grain size dynamic range increases from $10^{-3}$-0.1 $\rm\mu m$ to 0.1-10 $\rm\mu m$. Additionally, warmer gas suppresses low-mass star formation, and decreased opacities result in a greater proportion of gas in diffuse ionized structures.

\end{abstract}

\keywords{Star formation(1569)  --- Interstellar dynamics(839) --- Interstellar dust(836) --- Interstellar dust extinction(837) --- Interstellar dust processes(838)}

\section{Introduction} \label{sec:intro}

Dust plays a pivotal role in the processes involved in star formation and evolution of giant molecular clouds (GMCs). Dust absorbs stellar radiation in the far-ultraviolet (FUV) and re-emitting it in the infrared (IR) \citep{draine1984optical, mathis1990interstellar, li2001infrared, tielens:2005.book}. Additionally, dust significantly impacts the thermodynamics of GMCs \citep{mathis1983interstellar, draine:2003.dust.review}. As GMCs undergo collapse, developing high-density regions, dust becomes closely coupled to the gas through frequent collisions. These collisions facilitate the exchange of energy between dust and gas, resulting in the heating or cooling of dust grains and the opposite effect, cooling or heating, on the gas. Photoelectric heating, resulting from the absorption of radiation from the interstellar radiation field (ISRF) and neighboring stars, can also contribute to the heating of gas within the interstellar medium (ISM) \citep{goldreich1974molecular, leung1975radiation}. Additionally, dust grains act as efficient coolants, releasing energy through thermal emission and achieving a state of thermal equilibrium.

A critical property influencing the rates of the aforementioned processes is the size of dust grains, a parameter that is reasonably well-constrained within the diffuse ISM.  Canonical ISM grain models typically describe grains with an empirical Mathis-Rumpl-Nordsieck (MRN) spectrum with sizes up to $0.1 \rm \mu m$ \citep{mathis:1977.grain.sizes}. However, the regulation of the GSD involves various processes, including those inducing grain growth such as grain-grain coagulation \citep{chokshi1993dust} and accretion \citep{spitzerbook} , as well as grain destruction through thermal and non-thermal sputtering \citep{borkowski1995fragmentation,tielens1994physics}. Environmental conditions, including temperature, density, and turbulence, influence the rate of each dust grain process. Particularly, grain coagulation, the process that limits maximum grain size, is inefficient in the diffuse ISM but proceeds efficiently in the cool dense ISM \citep{yan2004dust}. This suggests that areas with increased density harbor larger grain sizes.

Therefore, it seems improbable that a single ISM grain size distribution (GSD) describes all star-forming environments across all galaxies throughout the history of the Universe. Indeed, observations of dense star-forming environments support this notion, revealing an abundance of larger dust grains \citep{johnson1964colors,savage1979observed,cardelli1988environmental,de2014extinction}. Furthermore, the ``coreshine'' effect, observed in the Mid-Infrared (MIR) and Near-Infrared (NIR) within dark clouds, can be ascribed to the presence of micron-sized grains that scatter background radiation \citep{pagani2010ubiquity, lefevre2014dust}. However, as reported by \citet{steinacker2015grain}, the GSD is not uniform across all clouds. Variations exist, with some clouds exhibiting sub-micron maximum grain sizes, while others have a potentially larger grain size cut-off, highlighting the diversity in grain populations. This diversity in the GSD extends to different galaxies as well \citep{pei1992interstellar, calzetti1994dust, salim2018dust, hopkins2004evolution, kriek2013dust}. Furthermore, recent simulations by \citet{hopkins:2021.dusty.winds.gmcs.rdis} demonstrated that dust dynamics alone can induce deviations from the typical MRN GSD within individual clouds or star-forming regions. Variations in the GSD are not confined to dense star-forming regions but are also observed across different sightlines within the diffuse Galactic ISM \citep{ysard2015dust, schlafly2016optical, wang:2017.extinction.law.variation.in.diffuse.ism}. Additionally, simulations by \citet{hirashita2009shattering, hirashita2023submillimetre} suggest that factors such as temperature, metallicity, and turbulence influence the maximum grain size, with sizes ranging from sub-micron to micron levels in star-forming environments.

The wide range of extinction curves and inferred GSDs across various spatial scales and environments underscores the importance of considering a range of grain sizes when studying physical processes within these regions. Dust grain properties profoundly affect the thermodynamics and evolution of molecular clouds, with grain size being a critical parameter.

At a fixed dust-to-gas ratio, smaller grains within a cloud could enhance dust shielding, potentially creating more favorable conditions for star formation. This correlation has been established and reported in simulations and observations by \citet{krumholz2008minimum, lada2010star, garcia2012star, gong2016molecular}. This heightened opacity would also increase the photoelectric heating rates, accompanied by higher collisional cooling rates due to the increased overall dust surface area.

However, it is important to note that the rates of photoelectric heating are contingent upon the incident FUV radiation, which diminishes with smaller grain sizes due to reduced photon penetration in the higher optical depth regime. Nevertheless, the potential increase in star formation could elevate the overall FUV radiation budget within the cloud. Additionally, changes in the dust cross section would impact other relevant processes, such as molecule formation rates. These interconnected processes are non-linear, making it uncertain a priori where the net effect would ultimately settle.


Therefore, to enhance our understanding of the intricate balance among photon penetration, heating rate, and the influence of grain size on these complex interactions, comprehensive investigations through rigorous simulations and theoretical models are imperative. This study introduces simulations of star formation that integrate detailed ISM physics, explicit dust dynamics, stellar formation, and feedback. The primary focus is to investigate the influence of grain properties, specifically grain size, on the thermodynamic characteristics of the clouds and how this parameter shapes the efficiency of star formation.

The paper is structured as follows: In Section \ref{sec:2}, we provide a concise description of the code and a description of the initial conditions for the runs. In Section \ref{sec:res}, we discuss the theoretical predictions of altering the GSD and compare them to the results obtained from our simulations. Finally, we conclude in Section \ref{sec:conc}.

\section{Simulations}
\label{sec:2}
\subsection{\SF simulation setup}

We utilize the \GIZMO code \citep{hopkins:gizmo} for conducting 3-dimensional radiation-dust-magnetohydrodnynamics (RDMHD) \SF \citep{starforge.fullphysics} simulations of star formation in giant molecular clouds, following the physics setup detailed in \citet{soliman2024dust}, which encompasses our complete \SF+dust physics modules. We offer a concise overview here; however, readers are encouraged to consult the aforementioned references as well as \citet{starforge.methods} for a more comprehensive description.

We utilise the \GIZMO Meshless Finite Mass magnetohydrodynamics (MHD) solver \citep{hopkins:gizmo.mhd, hopkins:gizmo.mhd.cg} for ideal MHD equations and the meshless frequency-integrated M1 solver for the time-dependent radiative transfer (RT) equations \citep{lupi:2017.gizmo.galaxy.form.methods, lupi:2018.h2.sfr.rhd.gizmo.methods, hopkins:2019.grudic.photon.momentum.rad.pressure.coupling, hopkins:radiation.methods, grudic:starforge.methods, hopkins.grudic:2018.rp, hopkins:2020.fire.rt}. The radiation is discretized into five frequency bands ($\lambda < 912 \angstrom$, $912<\lambda < 1550 \angstrom$, $1550<\lambda < 3600 \angstrom$, $3600<\lambda < 3\rm \mu m$ and $\lambda > 3\rm \mu m$), inducing processes such as photoionisation, photodissociation, photoelectric heating, and dust absorption, directly coupled with the dust distribution. Radiative cooling and heating terms also encompass metal lines, molecular lines, fine structure lines, and dust collisional processes, as detailed in \citet{fire3}. The rates for the dust radiative cooling and photoeletric heating, as well as other processes mentioned prior,  derive from interpolating the local dust particle distribution and local dust properties for each gas cell.

We assume an standard interstellar radiation field (ISRF) strength, based on solar neighborhood conditions \citep{draine2010physics}. Sink particles, which represent individual stars, are another source of radiation in the simulations. These particles form from gas cells that meet the criteria for runaway gravitational collapse, and follow the protostellar evolution model outlined by \citet{offner_protostar_mf}. As they grow through accretion, their luminosity and radius follow the \citet{tout_1996_mass_lum} relations, and they emit a black-body spectrum with an effective temperature $T_{\rm eff} = 5780 \, \text{K} \left({L_\star}/{R_\star^2}\right)^{1/4}$. They are also sources of protostellar jets, stellar winds, and potentially supernovae \citep{starforge.fullphysics}.

\subsection{Dust physics}

The dust physics we employ in our simulations mirrors the setup detailed in \citet{soliman2024dust}, where it is presented in greater detail. Furthermore, comprehensive studies of the modules and methods can be found in \citet{hopkins.2016:dust.gas.molecular.cloud.dynamics.sims, lee:dynamics.charged.dust.gmcs, moseley:2018.acoustic.rdi.sims, hopkins:2021.dusty.winds.gmcs.rdis, soliman2022dust}.

In our simulations, dust grains are modeled as ``super-particles'' using a Monte Carlo sampling technique \citep{carballido:2008.grain.streaming.instab.sims, johansen:2009.particle.clumping.metallicity.dependence, bai:2010.grain.streaming.vs.diskparams, pan:2011.grain.clustering.midstokes.sims, 2018MNRAS.478.2851M}. Each dust ``particle'' represents $N \gg 1$ dust grains with identical attributes, including grain size $\epsilon_{\text{grain}}$, mass $m_{\text{grain}}$, charge $q_{\text{grain}}$, and composition. The grain charge is determined for each particle for each timestep self-consistently by computing the collisional, photoelectric, and cosmic ray charging rates 
\citep{draine:1987.grain.charging, tielens:2005.book}. The grain sizes are statistically sampled to ensure that the ensemble of all particles adheres to an MRN size distribution with the desired dust-to-gas ratio, while also ensuring uniform particle distribution across logarithmic intervals in grain size. In particular, it is important to emphasise that the grain size for the individual particles, and thus the cloud average GSD does not evolve during the simulations; in other words, we do not model grain growth/coagulation or sputtering/destruction. However, the GSD within a particular volume can evolve over time as grains of varying sizes move in and out. Additionally, we do not include dust sublimation in our models, as this process typically becomes significant at temperatures exceeding 1500K. While this may overestimate dust opacity and dust cooling in warmer regions, it would minimally impacts the cooler, star-forming areas. Including sublimation would likely reinforce our conclusions by increasing the heating rates of warm gas.

We follow dust dynamics by accounting for drag, Lorentz, gravity, and radiation pressure forces. To ensure self-consistency, we interpolate local dust properties, such as the dust-to-gas (DTG) ratio and GSD, to their corresponding gas neighbors. Using this information, we compute local rates of various processes that dust is involved in, including:

\begin{itemize}
    \item \textbf{Radiative transfer}: Given our use of a simple MRN GSD and discretized radiation bins with constant dust opacity within each bin, we adopt a simplified model for dust absorption and scattering cross-sections. The dimensionless absorption+scattering efficiency is given by 
    \begin{equation}
        \langle Q (\grainsize, \lambda_{\rm eff})\rangle_{\rm ext} = \min \left( 2 \pi \grainsize/\lambda_{\rm eff}, 1 \right),
    \end{equation}

    where the effective wavelength is the geometric mean of the minimum and maximum wavelengths in the relevant range, $\lambda_{\rm eff} \equiv \sqrt{\lambda_{\min} \lambda_{\max}}$.

    \item \textbf{Collisional heating and cooling}: The dust  collisional cooling rate per unit volume, \(\Lambda_{\rm coll}\), is modelled as follows \citep{hollenbach1979molecule, hollenbach1989molecule, meijerink2005diagnostics}:
    \begin{align}
        \Lambda_{\rm coll} = &1.2 \times 10^{-32}\left(\frac{\dustgas}{0.01}\right)  (T - T_{\rm dust}) T^{1/2} \\&\left(1 - 0.8 e^{-75/T}\right) \left (\frac{100 \angstrom}{\grainsizemin} \right )^{1/2} \, \text{ergs} \, \text{cm}^3 \, \text{s}^{-1},
    \end{align}
    where $T$ and  $T_{\rm dust}$ are the temperatures of the gas and dust respectively, measured in Kelvin. 

    \textbf{Photoelectric heating}: The heating rate per unit volume due to the photoelectric effect on dust grains, $\Gamma_{\rm pe}$, is given by \citep{bakes1994photoelectric, wolfire2003neutral, wolfire_1995_fuv}:
\begin{align}
n\Gamma_{\rm pe} = 1.3 \times 10^{-24} n \epsilon G_0 \, \text{ergs} \, \text{cm}^{-3} \, \text{s}^{-1},    
\end{align}

where $n$ is the hydrogen number density in $\text{cm}^{-3}$ and $G_0$ is the FUV radiation field in Habing units. The heating efficiency, $\epsilon$, is defined as:
\begin{align}
\epsilon = & \frac{4.9 \times 10^{-2}}{1 + 4 \times 10^{-3} \left(G_0 T^{1/2}/2n_e\right)} \nonumber \\
& + \frac{3.7 \times 10^{-2} \left(T/10^4\right)^{0.7}}{1 + 2 \times 10^{-4} \left(G_0 T^{1/2}/2n_e\right)},
\end{align}
where $n_e$ is the electron number density. Note that polycyclic aromatic hydrocarbons (PAHs) are not included in our simulations, so their contribution to the photoelectric heating effect is excluded. The reasoning behind this choice and its implications are elaborated upon in Section \ref{sec:caveats}.

    \item \textbf{Molecular Hydrogen Formation on Dust Surfaces}: The formation rate of molecular hydrogen on dust, $\dot{n}_{\rm H_2, dust}$, is given by \citep{hollenbach1979molecule, jura1974formation, gry2002h2, habart2004some, wakelam2017h2}:
    \begin{align}
    \dot{n}_{\rm H_2, dust} = \alpha_{\rm H_2}(T) Z \dustgas n n_{\rm H I},
    \end{align}
    
    $Z$ is the metallicity as a fraction of solar, and $n_{\rm H I}$ is the $\rm H I$ number density. The rate coefficient, $\alpha_{\rm H_2}$, is computed as:
    \begin{align}
    \alpha_{\rm H_2} = \frac{9.0 \times 10^{-18} T^{0.5}}{1 + 0.04 T^{1/2} + 0.002 T + 8 \times 10^{-6} T^2} \, \text{cm}^3 \, \text{s}^{-1}.
    \end{align}
\end{itemize}

By incorporating these processes, we can effectively capture the influence of a live dust population on the thermochemical behavior and dynamics of the cloud.

\subsection{Initial conditions}

\begin{table*}
\centering
\begin{tabular}{llllllr}
\hline
$M_{\rm cloud}$ [$\rm M_\odot$] & $R_{\rm cloud}$ [pc] & $\Delta m$ [$\rm M_\odot$] & $\grainsizemax$ [$\rm \mu m$] & $B \rm\,  [\rm \mu G]$&$\sigma\rm \,[km/s]$&  Notes\\
\hline
$2 \times 10^3$ & 3 & $10^{-2}$ & 0.1 & 2 & 1.9& fiducial run \\
& & $10^{-3}$ & 1.0 &2 & 1.9& $\times $ 10 larger grains \\
& & $10^{-3}$ & 10 &2 & 1.9& $\times 100$ larger grains \\
\hline
$2 \times 10^4$ & 10 & $10^{-3}$ & 0.1 & 2 & 3.2& fiducial run \\
& & $10^{-2}$ & 1.0 & 2 & 3.2 & $\times$ 10 larger grains \\
& & $10^{-2}$ & 10 &2 & 3.2 & $\times$100 larger grains \\
\hline
\hline
\label{table}
\end{tabular}

\caption{The initial conditions for the simulations used in this study. The columns include: {(\bf{1})} Cloud mass $M_{\rm cloud}$. {(\bf{2})} Cloud radius $R_{\rm cloud}$. {(\bf{3})} Mass resolution $\Delta m$. {(\bf{4})} Maximum grain size $\grainsizemax$. {(\bf{5})} Initial magnetic field strength,
{(\bf{6})} Initial 3D turbulent velocity dispersion, {(\bf{7})} Notes indicating the main variations from the fiducial run.}

\end{table*} 

\label{sec:initial}

Our simulation setup involves a uniform-density turbulent molecular cloud surrounded by a diffuse warm ambient medium confined within a periodic box, whose dimensions are 10 times greater than the cloud's radius. The ambient medium has a density approximately $10^{3}$ times lower than that of the cloud. The initial velocity distribution follows a Gaussian random field, characterized by an initial virial parameter $\alpha_{\rm turb} = 5 \sigma^2 R/(3 G M_{\rm cloud}) = 2$. The initial magnetic field is uniform, establishing a mass-to-flux ratio 4.2 times the critical value within the cloud \citep{Mouschovias_Spitzer_1976_magnetic_collapse}. 

Our study includes two clouds of different masses. The first cloud has a mass of $M_{\rm cloud}=2 \times 10^{3} \rm M_\odot$ and a radius of $R=3\rm pc$, with a mass resolution of $\Delta m \sim 10^{-3}$. Additionally, we include a larger cloud configuration with $M=2 \times 10^{4} \rm M_\odot$ and radius $R=10\rm pc$, with a resolution of $\Delta m \sim 10^{-2}$. To precisely capture the dynamics of dust, we employ a mass resolution for dust super-particles that is four times higher than that of the gas \citep{moseley:2018.acoustic.rdi.sims}. Furthermore, cells associated with protostellar jets and stellar winds have a mass resolution ten times higher than that of the typical gas cells.

The initial dust distribution samples follows a statistically uniform DTG ratio $\initvalupper{\dustden} = \dustgas \initvalupper{\gasden}$ with $\dustgas=0.01$ corresponding to galactic values. The grains have an internal density of $\internaldensity \sim 2.25 \text{g/cm}^3$, which falls between the typical densities of carbonaceous and silicate dust grains.  Each particle is initialized with velocity corresponding to its nearest gas cell. Recall that the distribution of grain sizes samples from the empirical power law model proposed by \citet{mathis:1977.grain.sizes}, characterized by a differential number density represented as $d n_{\rm d} / d \grainsize \propto \grainsize^{-3.5}$. Ideally, the evolution of the GSD would be modeled self-consistently, but this requires simulating micro-physical processes across parsec-sized regions, which is currently unfeasible. Therefore, we consider three GSDs with maximum grain sizes of \(\grainsize^{\rm max} = 0.1 \, \mu \rm m\), \(\grainsize^{\rm max} = 1 \, \mu \rm m\), and \(\grainsize^{\rm max} = 10 \, \mu \rm m\), each with a minimum grain size of \(\grainsizemin = 0.01 \, \grainsizemax\). This approach approximates the shift towards larger grains in cool dense regions where coagulation is efficient and shattering is minimal \citep{birnstiel2011dust, hirashita2023submillimetre}.

We provide a summary of the initial conditions for all simulations discussed in this work in Table \ref{table}.

\section{Results}

\begin{figure*}
    \centering
    \includegraphics[width=0.86\linewidth]{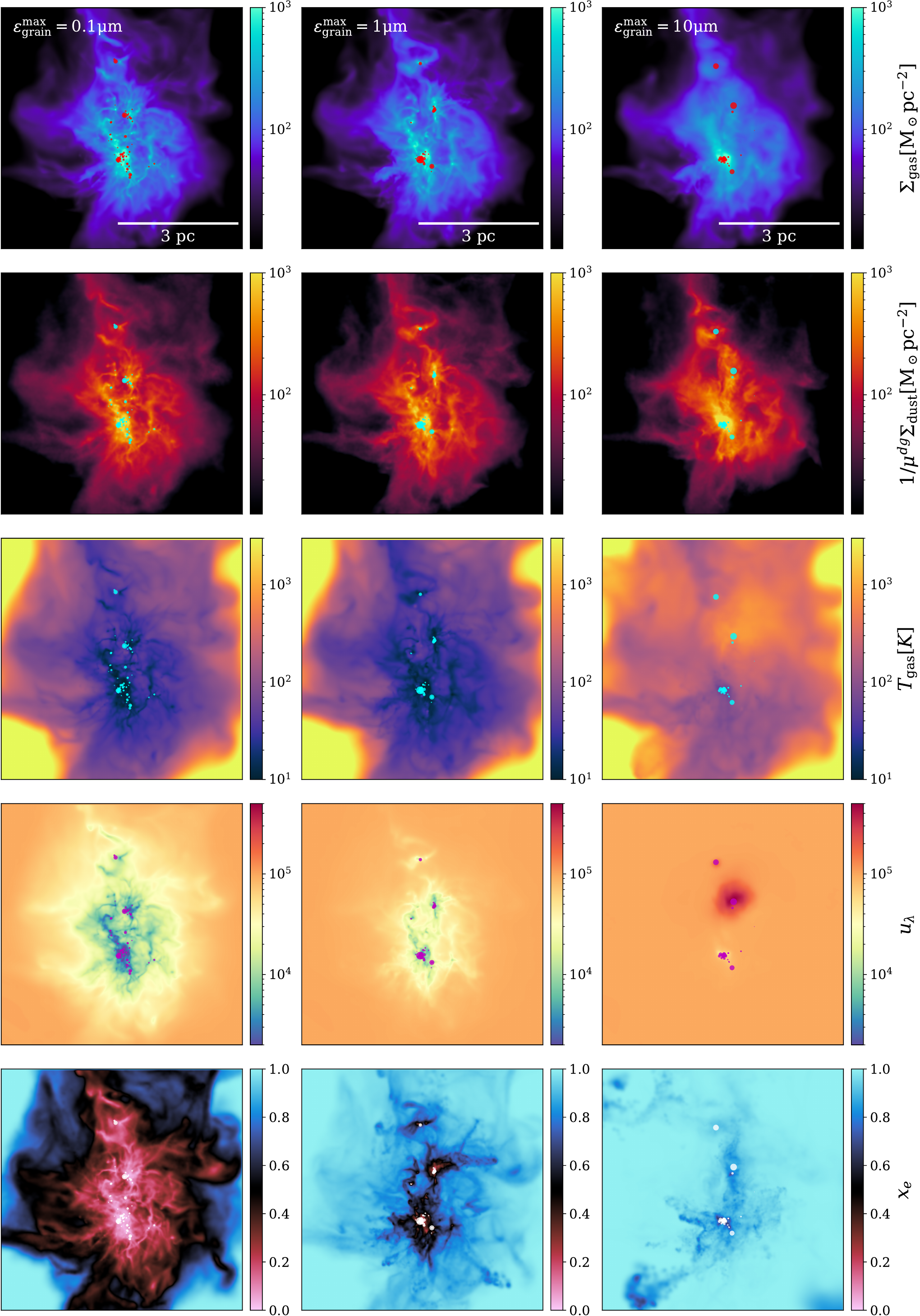}
    \caption{Morphological evolution of a $2\times 10^3 \rm M_{\odot}$ molecular cloud at $t\sim 3$ Myrs in simulations with varying grain sizes $\grainsizemax$. The top two rows show 2D integrated gas $\Sigma_{\rm gas}$ and dust $\Sigma_{\rm dust}$ surface densities, with stellar particles represented as circles, where their size corresponds to their stellar mass. Clouds with larger grain sizes exhibit more diffuse gas structures. However, this effect is less pronounced in the case of dust distribution. The third, fourth and fifth rows illustrate the projected gas mass-weighted mean temperature, mean radiation energy density of Far-UV/photoelectric band radiation ($912 \text{\r{A}} < \lambda < 1550 \text{\r{A}}$) in arbitrary units, and mean ionization fraction in the clouds. Clouds with larger grains exhibit increased temperatures, higher radiation energy densities, and consequently, elevated ionization fractions. }
    \label{fig:morph}
\end{figure*}

\label{sec:res}
\subsection{Theoretical expectations}

\label{expect}
Introducing variations in the GSD within the cloud can significantly affect its thermochemical properties. In the following section, we consider the expected changes resulting from these variations. Specifically, we examine how changes in the GSD would influence the optical depth $\tau_\lambda$.

For simplicity, we assume that the GSD remains constant along a line-of-sight through the cloud, and that the initial mean density within the cloud is spatially uniform. Recall that we model the distribution of dust particles according to an MRN size distribution, where $dn_{\text{d}}/d\grainsize = n_0\grainsize^{-3.5}$, with $n_0$ normalized to ensure $\int \grainmass dn_{\text{d}}/d\grainsize = \dustden^0 = \dustgas \gasden^0$. With these assumptions, we can express the optical depth $\tau_\lambda$ as:

\begin{align}
    \tau_\lambda = 2 R_{\rm cloud} \pi n_0\int^{\grainsizemax}_{\grainsizemin}  \grainsize^{-1.5} Q_{\rm abs} (\grainsize, \lambda)  \, d\grainsize,
\end{align}
where $Q_{\rm abs} (\grainsize, \lambda)$ is defined as follows 

\begin{align}
Q_{\rm abs} (\grainsize, \lambda)=
    \begin{cases}
    1 \quad &\lambda \leq 2 \pi\grainsize\\
    2\pi \grainsize/\lambda \quad & \lambda > 2 \pi\grainsize.\\
    \end{cases}
\end{align}

Therefore, given these assumptions and considering $\grainsizemax = 100 \grainsizemin$, the expression for $\tau_\lambda$ simplifies to

\begin{align}
\label{eq:tau}
\tau_\lambda=
    \begin{cases}
     15 \bar{\mu} R_{\rm cloud} \left ({\grainsizemax}\right )^{-1} \quad &\frac{\lambda}{2 \pi}  \leq \grainsizemin\\
     3\pi \bar{\mu} R_{\rm cloud}/ \lambda  \quad &\frac{\lambda}{2 \pi} \geq \grainsizemax\\
     5\sqrt{2 \pi /\lambda} \bar{\mu} R_{\rm cloud}\\ 
      \left ( 2 - \sqrt{\frac{2 \pi \grainsizemin}{\lambda \grainsizemax}} - \sqrt{\frac{\lambda}{2 \pi \left (\grainsizemax \right )^2}} \right ) \quad & \grainsizemin \leq \frac{\lambda}{2 \pi} \leq \grainsizemax, \\
    \end{cases}
\end{align}

where $\bar{\mu} \equiv\dustgas \gasden/ \internaldensity$.

While certain wavelengths of interest will fall within the intermediate regime ($\grainsizemin \leq \frac{\lambda}{2 \pi} \leq \grainsizemax$), we primarily focus on the geometric ($\lambda \leq 2 \pi \grainsize$) and Rayleigh ($\lambda \geq 2 \pi \grainsize$) regimes as they provide the most intuitive understanding. The intermediate regime mainly serves to interpolate between these two.

As highlighted in the expression above, modifying the GSD yields two distinct effects on dust opacity. Firstly, it dictates whether the majority of grains are situated in the geometric or Rayleigh regimes. Second, each of these regimes demonstrates a unique dependence on grain size: the Rayleigh regime maintains $\tau_\lambda$ independently of grain size, while in the geometric limit $\tau_\lambda \propto \left(\grainsizemax\right)^{-1}$. It is important to note that in this investigation, we explore variations in grain size while keeping the total dust mass constant. Consequently, increasing the grain size effectively reduces the total grain surface area, to which the geometric opacity is particularly sensitive. This explains why the Rayleigh opacity, being a bulk effect, does not exhibit any dependence on the grain size.

To identify the dominant opacity regime within the GMC, we examine the critical value of $\lambda/2 \pi$ where the transition between the geometric and Rayleigh regimes occurs in relation to grain size. 
Examining wavebands pertinent to star formation processes, specifically the FUV, Near Ultraviolet (NUV), Optical/NIR, and FIR bands tracked in our model, we note that these transitions occur at approximately $\grainsize \sim 10^{-2} \rm \mu m$, $0.05 \rm \mu m$, $0.1 \rm \mu m$, $10^{-2} \rm \mu m$, and $0.5 \rm \mu m$, respectively. Consequently, as $\grainsizemax$ increases from $0.1 \rm \mu m$ to $10 \rm \mu m$ (the distributions considered in this paper), a higher proportion of grains shift towards the geometric opacity regime. This shift is particularly pronounced at shorter wavelengths, such as in the UV band, where the opacity exhibits an inverse relationship with the maximum grain size. The FUV band opacity is particularly important for the thermodynamics of the GMC, as FUV photons play an important role in regulating the gas temperature through photoelectric dust heating.

Opacity-induced effects can drive highly nonlinear changes in cloud evolution. However, to first order, if the grain shielding dominates the clouds thermodynamics, larger grain sizes would enhance FUV radiation penetration, leading to warmer gas. This transition can significantly impact star formation rates and the properties of the stellar population. Warmer conditions, characterized by larger sonic scales and reduced density perturbations, would likely inhibit small-scale structure formation, giving rise to a smoother cloud morphology. Additionally, this would increase the Jeans mass, suggesting reduced low-mass star formation.

However, larger grain sizes also imply reduced photoelectric heating efficiencies. This reduction might, however, be counteracted by the the larger FUV flux, due to more photons penetrating, in addition to slower dust collisional cooling rates observed with larger grains ($\Lambda_{\rm coll} \propto \left(\epsilon_{\rm grain}^{\rm min}\right)^{-1/2}$). Ultimately, the interplay of these effects will dictate whether the gas experiences a net warming or cooling effect.

\subsection{Simulation results}

\subsubsection{Effects on cloud morphology}

\begin{figure*}
    \centering \includegraphics[width=0.95\linewidth]{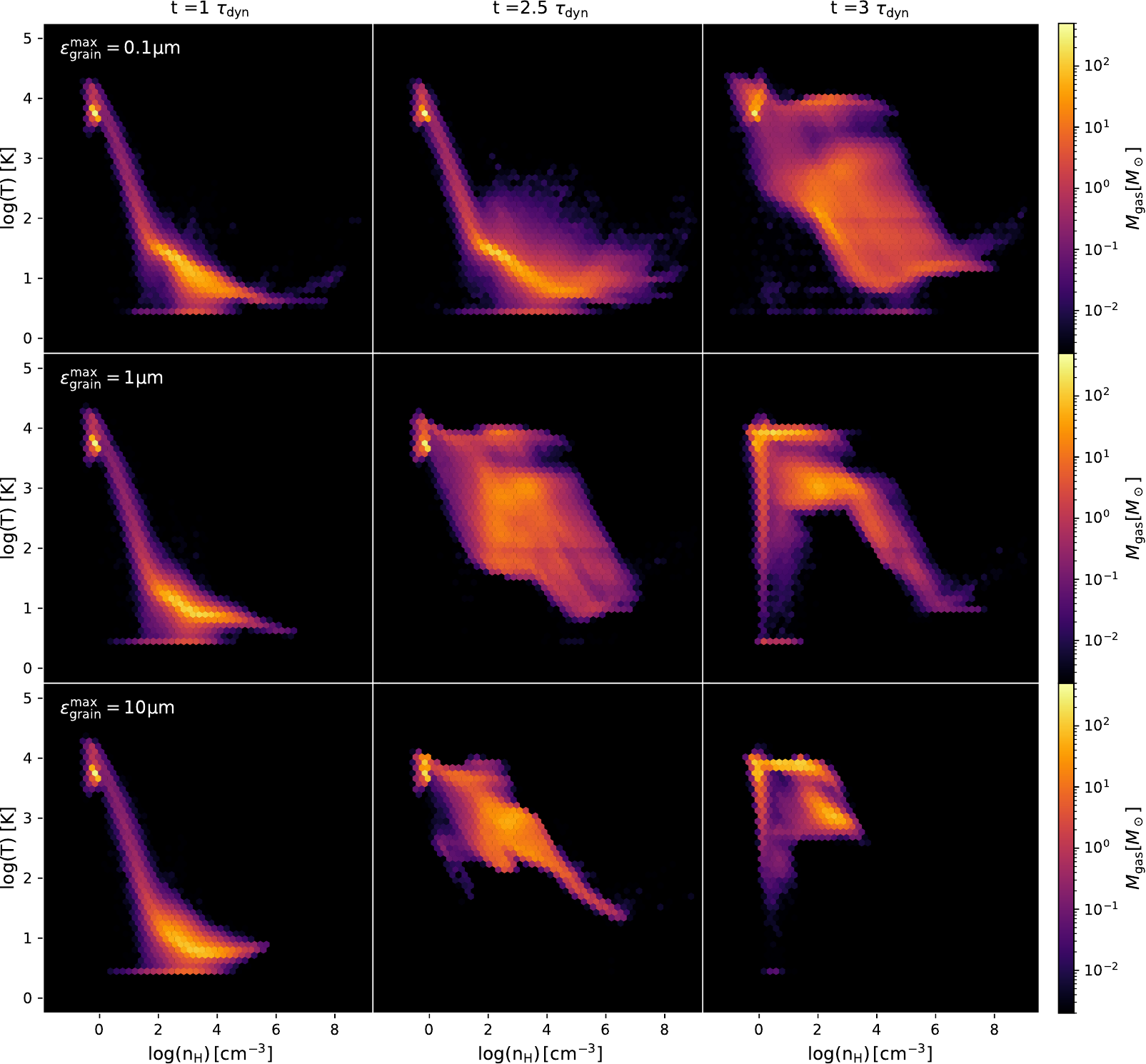}
    \caption{The temperature-density phase space diagrams showing the evolution of a cloud with a mass of $M_{\rm cloud} \sim 2 \times 10^3 \rm M_\odot$ at dynamical times $t\sim$ 1, 2.5, and 3 dynamical times. Different colors indicate the total gas mass within each state. At $t_{\rm dyn} \sim 1$, all clouds exhibit comparable states, with smaller grain clouds that extend to cooler and denser gas components. By $t \sim 2.5 t_{\rm dyn}$, star formation concludes, and larger grains exhibit higher average temperatures. At $t \sim 3 t_{\rm dyn}$, gas with $n_{\rm H}\leq 10^4 \rm cm^{-3}$ is predominantly hot ($T\sim 10^3$K), with denser gas being cooler. The $\grainsizemax = 1 \rm \mu m$ component lacks a dense counterpart, remaining mostly hot with an average temperature of $T\sim 10^3$K. Note that the clustering at low density density is an artifact of the diffuse ambient medium within the simulation box. Additionally, since all material is confined within a finite box, this prevents the gas from becoming more diffuse. Likewise, although to a lesser extent as less gas resides at low temperatures, the clustering at low temperatures is attributed to our temperature floor set at 2.73 K. }
    \label{fig:phase}
\end{figure*}

In this study, we conducted simulations of GMCs with initial conditions detailed in Section \ref{sec:initial}. We systematically varied the maximum grain size while maintaining a fixed DTG ratio at the start of each simulation. Building upon the theoretical framework outlined in the previous section, this section presents the results obtained from our simulations.

In Figure \ref{fig:morph}, we present the morphology of molecular clouds, each with an initial mass of approximately $2\times 10^3 \rm M_{\odot}$ and a resolution of $\Delta m \sim 10^{-3} \rm M_\odot$, evolved for $\sim$3 Myrs. From left to right, the columns represent clouds with different maximum grain sizes: $\grainsizemax = 0.1 \rm \mu m$, $\grainsizemax = 1 \rm \mu m$, and $\grainsizemax = 10 \rm \mu m$ respectively. The top two rows provide a visual representation of the 2D integrated gas $\Sigma_{\rm gas}$ and dust $\Sigma_{\rm dust}$ surface densities, with stellar particles shown as circles, scaled according to their masses. The subsequent rows describe the thermodynamic and radiative properties of the cloud. The third row shows the average temperature of the gas, while the fourth row shows the average radiation energy density associated with FUV radiation in the range $912 \text{\r{A}} < \lambda < 1550 \text{\r{A}}$ in arbitrary units. The fifth row displays the gas mass-weighted ionization fraction of the gas.

The clouds with larger grain sizes exhibit higher FUV radiation energy densities relative to their smaller grains counterparts. This is due to reduced FUV opacity, as demonstrated in Equation \ref{eq:tau}, allowing radiation to propagate more extensively throughout the cloud. Consequently, this leads to elevated temperatures driven by the photoelectric effect on dust grains.

In line with our predictions outlined in Section \ref{expect}, the increase in temperature is accompanied by a reduction in small-scale structures and weaker density fluctuations. This effect is particularly evident in the diffuse gas structures formed in our $\grainsizemax = 10 \rm \mu m$ simulation. In contrast, a similar smoothing effect is not observed in the dust structure. This discrepancy is to be expected, as the dust’s thermal velocity dispersion is lower than that of the gas, making it less affected by higher temperatures. However, the dust structures would still experiences some broadening due to its coupling with the gas dynamics. However, despite the consistent difference in FUV radiation energy density across the different grain size runs in the early stages, a significant temperature increase is observed only when the most massive stars form and emit substantial amounts of radiation. Specifically, a $\sim 4 M_\odot$ sink particle coincides with the high radiation energy density and temperature peaks. The radiative feedback from this star warms the gas in the cloud, facilitating the smoothing of the gas structure within $\leq 0.1$ Myr.

Furthermore, there is a distinct contrast in the ionisation fraction among different grain size runs. In the simulation with $\grainsizemax=0.1 \rm \mu m$, the majority of the cloud remains predominantly molecular. However, increasing the grain size by a factor of 10 confines molecular regions to the dense central core, while the majority of the cloud is in a predominantly ionized state. A further increase by a factor of 10 results in an almost fully ionized cloud. This marked difference is expected given the heightened radiation energy density and elevated temperatures in clouds with larger grains. In addition, the rate of molecular hydrogen formation decreases as the total surface area-to-mass ratio of the grains decreases.

\begin{figure*}
    \centering \includegraphics[width=0.95\linewidth]{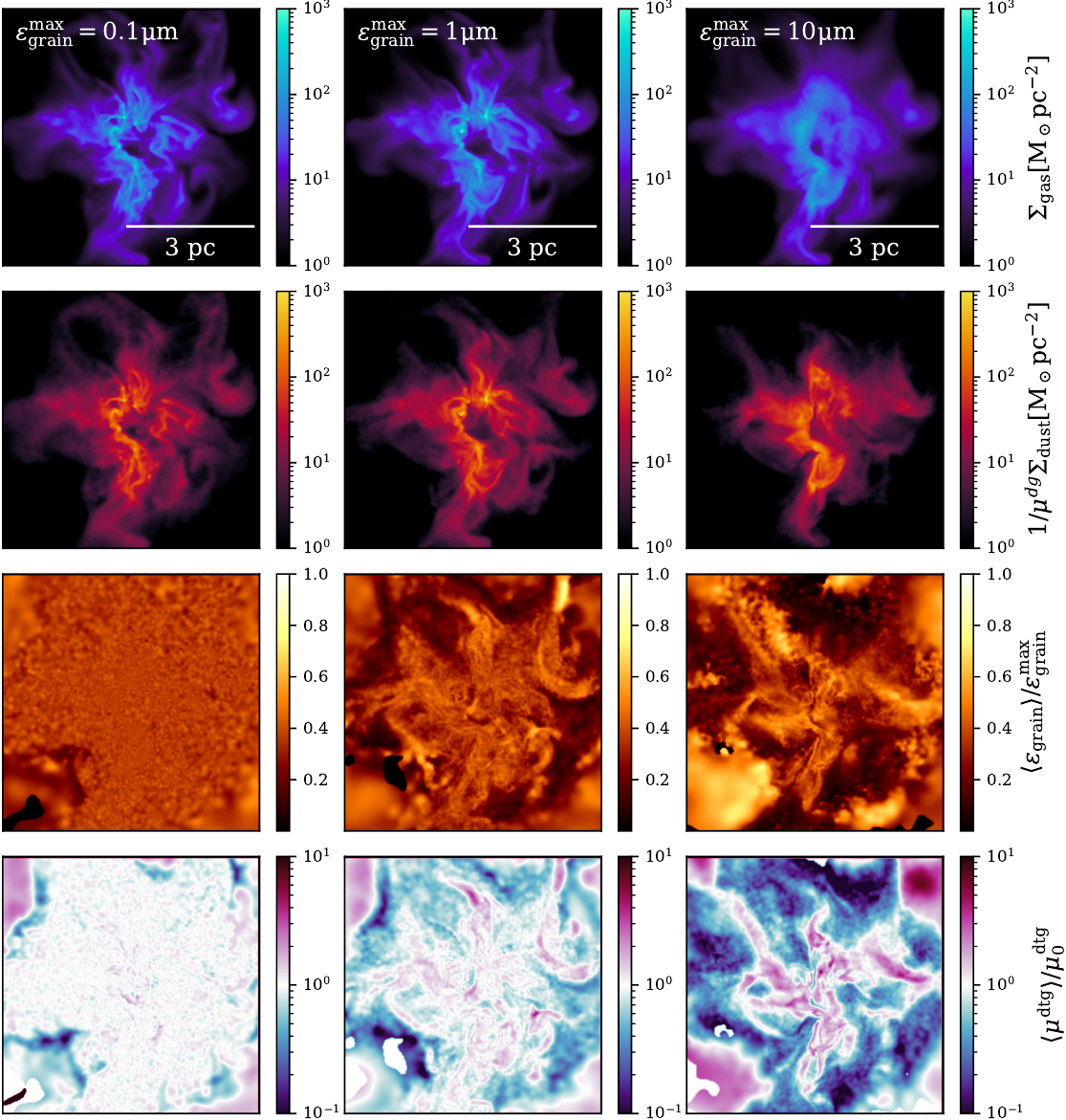}
    \caption{Morphological features of a 0.5 pc thick cross-section through a molecular cloud with a mass of $2\times 10^3 \rm M_{\odot}$ at $t\sim 3$ Myrs, simulated with varying grain sizes $\grainsizemax$. From top to bottom: 2D integrated gas $\Sigma_{\rm gas}$ and dust $\Sigma_{\rm dust}$ surface densities, normalized mean grain size, and normalized mean dust-to-gas (DTG) ratio $\dustgas$. Note that we present $1 / (\dustgas \Sigma_{\rm dust})$ for ease of comparison. The cloud with $\grainsizemax =0.1 \rm \mu m$ shows negligible fluctuations in grain size spatial distribution and DTG ratios, with dust closely following the gas distribution. Conversely, clouds with larger grains exhibit fluctuations in grain size spatial distribution and DTG ratios. In the case of $\grainsizemax=10 \rm \mu m$, this results in order of magnitude fluctuations in the DTG ratio, particularly in regions dominated by large grains. }
    \label{fig:grainsize}
\end{figure*}

To further explore the thermodynamic evolution of the three clouds, Figure \ref{fig:phase} illustrates the temperature-density phase-space diagram at different dynamical times $t_{\rm dyn}$. The three panels represent time intervals of $t_{\rm dyn} \sim 1$, $2.5$, and $3$, respectively, with the colour map corresponding to the total gas mass within a given state.

At the initial stage ($t_{\rm dyn} \sim 1$), all three clouds exhibit similar distributions. However, the clouds with smaller grain sizes contain more gas in cool ($T \lesssim 20$ K) and dense structures (number densities of $n_{\rm H} \gtrsim 10^6 \rm cm^{-3}$), although this represents only a small fraction of the total gas. As a result, these regions have a higher number of cores prone to gravitational collapse, leading to an earlier onset of star formation. We point out that the gas component at $T\sim 10^4$ K and $n_{\rm H} \sim 1 , \rm cm^{-3}$ corresponds to the hot gas bath that surrounds the molecular cloud as per our initial conditions.

By $t_{\rm dyn} \sim 2.5$, star formation has proceeded to completion in all three clouds. However, clouds with larger grain sizes contain warmer gas due to weaker shielding from these larger grains. This shift to warmer gas occurs only after most stars have formed, enhancing the radiation field and leading to a rapid transition to a warm, quenched cloud. Prior to this, the three clouds with different grain sizes have similar temperatures and appear comparable. The temporal evolution of stellar mass and temperature is shown in Figure \ref{fig:test2}, which we discuss in the following subsection.

Moving to $t_{\rm dyn} \sim 3$, most gas with $n_{\rm H} \leq \sim 10^4 , \rm cm^{-3}$ is predominantly warm ($T\sim 10^3$ K), while denser gas remains fairly cool. Notably, the component with $\grainsizemax = 1 , \rm \mu m$ lacks a dense counterpart and is mostly warm, with an average temperature of $T\sim 10^3$ K. This component is ionized, corresponding to the warm ionized medium (WIM) with no cold neutral medium (CNM) component.

In Figure \ref{fig:grainsize}, we present the morphology of a 0.5 pc thick slice through our $2\times 10^3 \rm M_{\odot}$ cloud at $t\sim 3$ Myrs evolved with different GSDs. The top two rows present 2D integrated surface densities of gas $\Sigma_{\rm gas}$ and dust $\Sigma_{\rm dust}$, while the third row displays the average grain size across the slice normalized to $\grainsizemax$ in the cloud. The fourth row shows the mean DTG ratio with respect to the cloud's mean value.

In the $\grainsizemax = 0.1 \rm \mu m$ cloud, the gas and dust are well-coupled, evident in their closely aligned spatial distributions. This implies that the grain sizes within the simulation predominantly have stopping times $\ts$ much shorter than the timescale of gas dynamics. As a result, there are no discernible variations observed in the DTG ratio, and there is a uniform distribution of grains across all sizes.

In the $\grainsizemax=1 \rm \mu m$ cloud, larger grains are less effectively coupled to the gas, therefore they do not trace the gas dynamics as well as their smaller grain counterparts. This discrepancy introduces fluctuations in the spatial distribution of grain sizes, with larger grains lagging behind the gas flow. This effect is particularly pronounced in regions of low density where large grains would encounter even lengthier stopping times as $\ts \propto \grainsize/\gasden$. Given that larger grains contribute substantially to the overall dust mass under an MRN GSD, this poor dust coupling for large grains can drive large-scale fluctuations in DTG ratios. The most striking results emerge in the $\grainsizemax=10 \rm \mu m$ cloud, where even more pronounced fluctuations in GSD are observed. This leads to order of magnitude variations in the DTG ratio, particularly in regions with predominantly large grains.

\subsubsection{Effects on initial mass function}

\begin{figure*}
    \centering
    \includegraphics[width=1\linewidth]{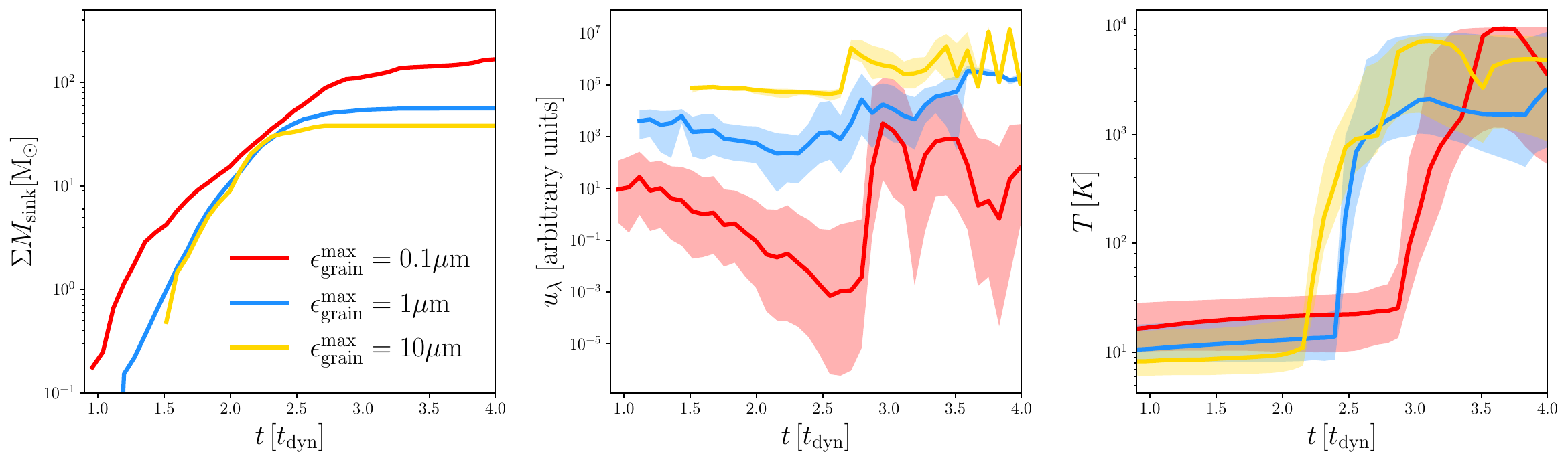}
    \caption{Evolution of star-forming Giant Molecular Clouds (GMCs) over time, represented in units of the dynamical time of the cloud. Left panel: Total stellar mass formed. Middle panel: Average photon energy density within 0.1 parsec spherical regions around the formed stars. Right panel: Average gas temperatures outside the 0.1 parsec regions. The shaded areas represent a range of one standard deviation. The simulations compare GMCs with an initial cloud mass of $M_{\rm cloud} \sim 2 \times 10^{3} \rm M_{\odot}$ and different grain-sizes with maximum grain-sizes of $\epsilon^{\rm max}_{\rm grain}=$ 0.1 $\rm \mu$m (blue), 1 $\rm \mu$m (yellow), and 10 $\rm \mu$m (red). Larger grains lead to lower star formation efficiency, with roughly a tenfold increase in total stellar mass observed for the 0.1 $\mu$m grains, exhibiting a $\sim$10\% star formation efficiency compared to the 10 $\mu$m runs with $\sim 1\%$. Smaller grains provide stronger dust shielding, resulting in a cooler gas that is more prone to gravitational collapse and star formation. }
    \label{fig:test2}
\end{figure*}

We quantify the impact of larger dust grains on star formation in Figure \ref{fig:test2}, which illustrates the evolution of star-forming GMCs over time, represented in units of the cloud's dynamical time.  The right panel shows the total stellar mass formed, the middle panel displays the mean FUV band radiation energy density within 0.1 parsec spherical regions around the formed stars, and the left panel shows the average gas temperatures outside these regions.

The $\epsilon_{\rm grain}^{\rm max} = 0.1\mu$m simulation exhibits an earlier onset of star formation compared to $\epsilon_{\rm grain}^{\rm max} = 1\mu$m and $\epsilon_{\rm grain}^{\rm max} = 10\mu$m simulations. However, the difference in the timing of star formation onset is minimal. Initially, the gas is only radiated by the relatively faint ISRF. This diminishes the importance of factors that dependent on grain size such as shielding effects. Additionally, variations in the GSD result in relatively negligible differences in the net dust-mediated cooling and heating rates, leading to minimal impact on the overall thermal state of the gas during the early stages, before a strong radiation field is established. As a result, the gas sustains comparable average temperatures across various grain sizes. This remains the case until enough sinks form, particularly massive sinks, and begin to contribute to the radiation field. At approximately 1.8 dynamical times, all clouds exhibit similar average temperatures and attain the same stellar mass. However, smaller grains offer higher dust opacity, which reduces the propagation of UV flux from the stars to the surrounding gas. As a result, the clouds with smaller grains maintain a cooler temperature for a longer duration.

As assumed by our opacity toy model, the UV radiation energy density scales inversely with the square of the grain size. This enhanced dust shielding enables the cloud to sustain ongoing star formation until the gas eventually heats up, and stellar feedback leads to the evacuation of the cloud and halts further star formation. A significant contrast in the star formation efficiency is evident when comparing clouds with the largest grains to those with the smallest grains. Specifically, the cloud with 10 $\rm \mu m$ grains converts only 1\% of its mass into stars, whereas the cloud with 0.1 $\rm \mu m$ grains exhibits roughly a tenfold higher star formation efficiency, converting 10\% of its mass into stars.

\begin{figure}
    \centering \includegraphics[width=0.95\linewidth]{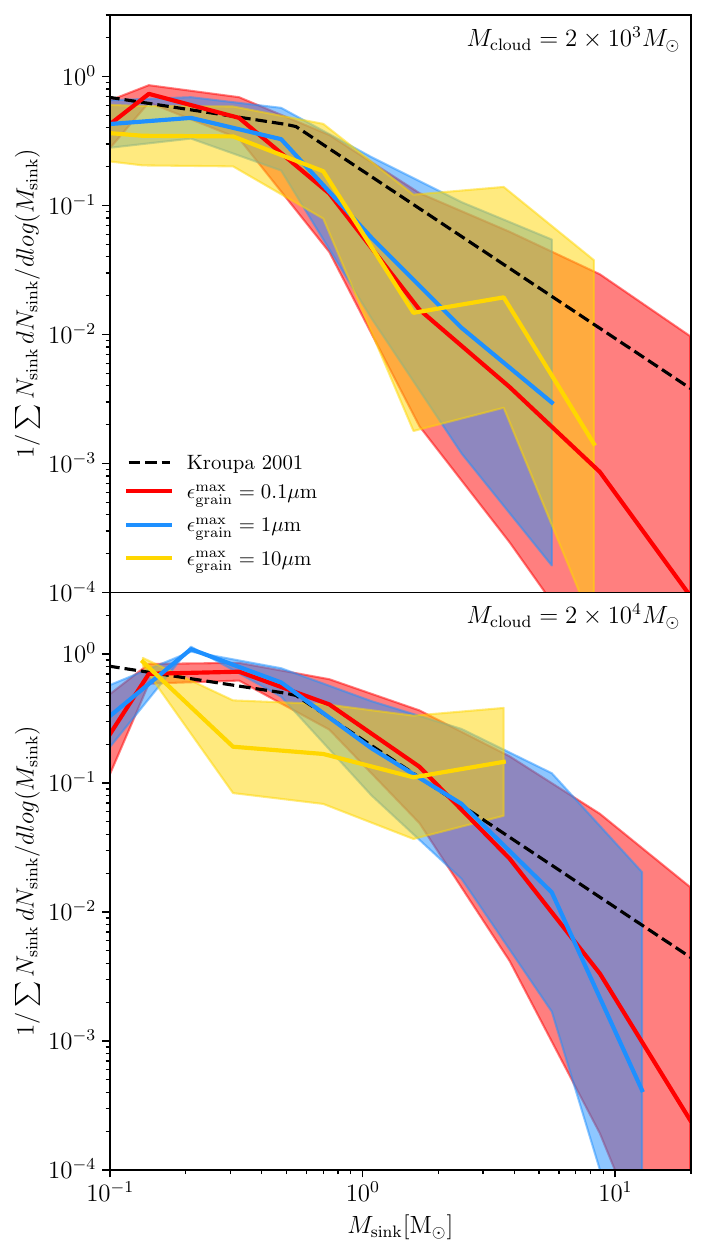}
    \caption{The mass distribution of sink particle masses per logarithmic mass interval, normalized to the total number of sinks compared to a \citet{kroupa2001variation} inital mass function. The shaded region shows the Poisson sampling error. Top: Distribution for our $M_{\rm cloud} \sim 2\times 10^3 \rm M_\odot$ cloud at $\Delta m \sim 10^{-3} \rm M_\odot$ resolution . Bottom: Distribution for our $M_{\rm cloud} \sim 2\times 10^4 \rm M_\odot$ at $\Delta m \sim 10^{-2} \rm M_\odot$. Larger grains, particularly in the clouds with larger masses, lead to a more restricted range of stellar masses. Nonetheless, we do not observe any significant changes in the distribution driven by variations in grain size. }
    \label{fig:imf}
\end{figure}

In Figure \ref{fig:imf} we present the mass function of sink particles that form in clouds with $M_{\rm cloud} \sim 2\times 10^3 \rm M_\odot$ (top) and $M_{\rm cloud} \sim 2\times 10^4 \rm M_\odot$ (bottom). The results for different grain sizes are compared to the initial mass function from \citet{kroupa2001variation}. The shaded region represents the Poisson error. We find that clouds with larger grains yield stellar populations with a narrower mass range. Specifically, as the grain size increases by a factor of 10, the range of sink masses decreases by a factor of 2. The higher minimum sink mass can be attributed to the higher expected jeans mass, while the high-mass trucation is likely due to the reduced number of sinks forming in the large grain runs, resulting in limited sampling at the higher mass end. While our simulations indicate a higher mean sink mass in setups with larger grains, the reduced sink count, especially at the high mass end due to sampling, and sparse statistical data highlight the necessity for a more comprehensive statistical analysis to attain a precise understanding of the distribution. Presently, our simulations tentatively suggest that grain size has minimal impact on the mass function, within statistical error margins.

Another trend in Figure \ref{fig:imf} is the reduction in low-mass sink particles ($M_{\rm sink} \leq 1 , \rm M_\odot$) for clouds with larger grains. In the smaller cloud, the fraction of low-mass sinks decreases from $\sim $ 0.8 to about $\sim $ 0.6, and in the larger cloud, it drops from $\sim $ 0.6 to around $\sim $0.2 as the grain size increases from 0.1 $\mu$m to 10 $\mu$m. This decline is attributed to the elevated temperature, which leads to a larger Jeans mass, thus inhibiting smaller cores from meeting the criteria for collapse in warmer environments. This sequence of events underscores the intricate interplay among grain size, radiation, temperature, and structural characteristics in shaping the dynamic evolution of molecular clouds.

\subsection{Caveats}
\label{sec:caveats}
Several caveats should be considered when interpreting our findings. Firstly, our study does not encompass the full spectrum of factors that may affect dust opacities within star-forming regions. Specifically, variations in the DTG ratio due to non-power law characteristics in the GSD and grain chemistry are not explored. Additionally, we use a highly simplistic toy model for the dust opacities. This is intentional as it is designed to capture the leading-order physics while simplifying the complex interplay of non-linear effects. Recall that we assume a constant \(Q_\text{ext}\) across the 990-1550 $\angstrom$ waveband and model \(Q \propto w \pi \epsilon / \lambda\), with \(\lambda\) as the geometric mean of the waveband limits. Compared to detailed models like \citet{zubko2004interstellar}, our simplified approach underestimates \(Q_\text{ext}\) at shorter wavelengths and overestimates it at longer wavelengths for grains smaller than \(10^{-2} \ \mu\text{m}\), while for larger grains, \(Q_\text{ext}\) remains \(\propto 1/\lambda\). However, these deviations stay within a factor of $\sim 2$ and largely average out over the waveband range. Similarly, deviations from $Q_{\text{ext}} \propto \epsilon$ remain within a factor of two across four orders of magnitude in grain size, indicating that our conclusions on grain size effects on opacity and thermodynamics are qualitatively robust even with more detailed models.

While our study provides insights into the effects of varying initial GSDs on star formation, the absence of these additional factors may limit the comprehensiveness of our conclusions. All else equal, and considering the impact of these factors on dust opacity, their incorporation would effectively lead to a re-scaling of opacities for a given range in grain size. Consequently, we anticipate that, to leading order, these factors would produce similar effects on star formation as those observed in our study.

In addition, we do not incorporate the role of PAHs, as it falls beyond the current scope of our investigation. PAHs behave as gas-phase molecules, requiring distinct physics to accurately model their effects and behavior within this environment. Our paper primarily focuses on elucidating the first-order effects of large versus small grains. Consequently, we have omitted the effects of PAHs from our analysis. Nevertheless, considering PAHs would likely amplify the effects we report. Assuming, the main finding of increased FUV radiation due to lower dust grain opacities remains robust with the inclusion of PAHs, their presence would likely enhance photoelectric heating for a given radiation field. This enhancement could lead to earlier or higher temperature increases and ionization fractions in the gas. The respective effects of PAHs and grains on radiative transfer and thermochemsitry with GMCs, is complex. The outcomes of including another dust species that plays an important role in these process might yield effects different from those discussed in the current study,  emphasizing the need for further investigation in subsequent studies.

Secondly, while we systematically explore the effects of different initial GSDs, the evolution of GSDs over time within actual star-forming regions is influenced by a multitude of dynamic factors. Processes such as grain destruction, coagulation, and growth exhibit variations across different environments, introducing complexities that are not fully captured in our model. In future work, we aim to model the evolution of GSDs within star-forming regions, following a similar approach as demonstrated in recent work by \citet{choban2022galactic}. This endeavour will contribute to a more comprehensive understanding of the interplay between dust properties and star formation processes.

\section{Conclusions}
\label{sec:conc}

The dust GSD likely varies significantly across different star-forming regions, with evidence of variability observed in diverse galaxy extinction curves. Observational data from dust scattering in dark clouds and molecular cloud extinction curves further suggests deviations from the canonical diffuse ISM GSD, indicating larger grain sizes in these regions. This variability may be more pronounced in high-redshift systems and across various star-forming environments.

In this study, we conducted a series of RDMHD \GIZMO simulations focusing on star-forming GMCs with different GSD, specifically with $\grainsizemax = 0.1, 1, 10 \rm \mu m$. 

Our simulations included various competing effects, such as dust collisional heating/cooling, photoelectric heating, and dust shielding, our results indicate that larger grain sizes lead to a decrease in star formation efficiency. Our findings highlight that the rate of star formation declines more rapidly in clouds with larger grains. The decrease in star formation efficiency is due to enhanced radiation penetration through the cloud, facilitated by reduced dust shielding. This results in more efficient heating and ionization, all of which are non-linear processes.

In summary, the observed effects emphasize the necessity of a careful consideration of grain size variations when interpreting and modeling the physical processes within star-forming regions.
\vspace{1em}

\noindent Support for for NS and PFH was provided by NSF Research Grants 1911233, 20009234, 2108318, NSF CAREER grant 1455342, NASA grants 80NSSC18K0562, HST-AR-15800. Support for MYG was provided by NASA through the NASA Hubble Fellowship grant \#HST-HF2-51479 awarded  by  the  Space  Telescope  Science  Institute,  which  is  operated  by  the   Association  of  Universities  for  Research  in  Astronomy,  Inc.,  for  NASA,  under  contract NAS5-26555. Numerical calculations were run on the TACC compute cluster ``Frontera,'' allocations AST21010, AST20016, and AST21002 supported by the NSF and TACC, and NASA HEC SMD-16-7592.  This research is part of the Frontera computing project at the Texas Advanced Computing Center. Frontera is made possible by National Science Foundation award OAC-1818253.

\datastatement{The data supporting this article are available on reasonable request to the corresponding author.}

\software{\href{https://matplotlib.org/}{\fontfamily{cmtt} \selectfont matplotlib} \citep{Hunter:2007}, \href{https://numpy.org/}{\fontfamily{cmtt} \selectfont NumPy} \citep{harris2020array}, \href{https://scipy.org/}{ \fontfamily{cmtt} \selectfont SciPy} \citep{2020SciPy-NMeth}, \href{https://cmasher.readthedocs.io/index.html}{\fontfamily{cmtt} \selectfont CMasher} \citep{cmasher}.}

\bibliography{sample631}{}
\bibliographystyle{aasjournal}

\end{document}